\documentclass[11pt]{article}

\usepackage[utf8]{inputenc}
\usepackage[T1]{fontenc}
\usepackage{amsmath}
\usepackage{amssymb}
\usepackage{amsfonts}
\usepackage{amsthm}

\usepackage{geometry}
\usepackage{setspace}
\usepackage{graphicx}
\usepackage{caption}
\usepackage{subcaption}
\usepackage{booktabs}
\usepackage{array}
\usepackage{tabularx}
\usepackage{tikz}
\usetikzlibrary{fit,backgrounds,decorations.pathreplacing,calligraphy,positioning}
\usepackage{algorithm}
\usepackage{algorithmic}

\usepackage{xurl}
\usepackage{hyperref}
\hypersetup{
    colorlinks=true,
    linkcolor=blue,
    citecolor=blue,
    urlcolor=blue
}

\usepackage{enumitem}
\usepackage{xcolor}
\usepackage[style=numeric-comp,sorting=none]{biblatex}
\usepackage{tikz}
\usetikzlibrary{arrows.meta,positioning}

\newcommand{\ket}[1]{|#1\rangle}

\newcommand{\Aprep}{\mathcal{A}}

\usepackage{authblk}

\title{%
  \textbf{Deterministic Quantum Search for NP Problems Through Verifier-Based Oracle Construction}\\[4pt]
 
}

\author[1]{Soumyajyoti Chatterjee\thanks{\texttt{soumyajyoti@cdac.in}}}
\author[1]{Harishankar Mishra\thanks{\texttt{harishankarm@cdac.in}}}
\author[1]{Asvija Balasubramanyam\thanks{\texttt{asvijab@cdac.in}}}
\author[2]{S D Sudarsan\thanks{\texttt{sdd@cdac.in}}}
\author[1]{Gudapati Naresh Raghava\thanks{\texttt{nareshraghava@cdac.in}}}
\affil[1]{Quantum Technology Group, Centre for Development of Advanced Computing (C-DAC), Bangalore}
\affil[2]{Centre for Development of Advanced Computing (C-DAC), Bangalore}

\providecommand{\keywords}[1]{%
  \par\medskip\noindent\textbf{Keywords: }#1\par}
\begin{document}

\maketitle

\begin{abstract}

Problems in the class NP are important in computational science, where a given candidate solution can be verified in polynomial time, while finding a solution may require evaluating the entire search space in the worst case. Classical exhaustive approaches generally solve such problems by evaluating the verifier function over all possible inputs, resulting in a worst-case query complexity of \(O(N)\), where \(N=d^n\) represents the size of the search space, \(d\) is the number of possible values for each variable, and \(n\) is the number of variables. Quantum search algorithms can solve unstructured search problems more efficiently by reducing the query complexity to \(O(\sqrt{N})\), providing a quadratic speedup over classical exhaustive search. However, conventional quantum search formulations typically assume access to an oracle that identifies valid solutions, while standard implementations generally yield probabilistic success. Constructing such an oracle directly from a classical verifier and achieving deterministic identification present additional challenges. In this work, we propose a generalised quantum oracle construction method for NP problems, together with a deterministic quantum search framework that identifies satisfying assignments using their corresponding classical verifier functions. The proposed framework retains the quadratic query speedup of quantum search over classical exhaustive search. We further establish that the constructed oracle can be implemented using a number of gates polynomial in the size ($L$) of the binary encoding of verification function, consistent with the polynomial-time evaluation of the corresponding classical verifier, while the total qubit requirement remains $O(\mathrm{poly}(L))$. The proposed construction is demonstrated for multiple \(N\)-SAT instances, where circuit-level simulation confirms deterministic identification of the satisfying assignments. These results demonstrate that deterministic identification of valid solutions can be achieved for problems in the class NP while retaining the quadratic query advantage of quantum search over classical exhaustive search.


\end{abstract}



%
\section{Introduction}
\label{sec:introduction}

Problems in the class NP form an important category of computational
problems in which a proposed solution can be verified in polynomial time, while finding such a solution is not known to admit an efficient polynomial-time procedure for all the problems in this class \cite{sipser2012theory}. This distinction between verification and solution finding arises in several
important computational problems, including satisfiability \cite{cook1971complexity, levin1973universal},
Hamiltonian Cycle, Clique, and Vertex Cover \cite{karp1972reducibility},
Subset Sum \cite{garey1979computers}, several decision versions of scheduling
\cite{lenstra1978complexity}, and the decision version of the Traveling Salesman
Problem \cite{garey1979computers}, where a candidate solution can be verified
efficiently while finding a solution satisfying the required constraints may
be computationally difficult. Thus, a general NP problem can be viewed as a search for an input that satisfies a polynomial-time verifier. Although the verifier can efficiently determine whether a given input is a valid solution, the absence of a known efficient procedure for finding such an input makes the search computationally challenging.

The conventional classical exhaustive approach for solving such search problems is to examine the possible inputs individually and evaluate each input using the corresponding verifier. For an input space of size $N=d^n$, classical exhaustive search may require evaluating all $N$ possible inputs in the worst case, resulting in a query complexity of $O(N)$. 


Quantum computing provides an alternative approach to such search problems by allowing candidate inputs to be represented simultaneously in a quantum superposition. Grover's algorithm demonstrated that an unstructured search over a space of $N$ elements can be performed using $O(\sqrt{N})$ oracle queries, providing a quadratic improvement over the classical search complexity \cite{grover1996fast,grover1997quantum}. The tight bounds on quantum searching established by \cite{boyer1998tight} further characterize this quadratic advantage for arbitrary numbers of target states. The amplitude amplification framework subsequently generalized the basic Grover procedure to arbitrary initial state preparations, while retaining the quadratic improvement in query complexity \cite{brassard2002quantum}. However, these quantum search procedures are generally probabilistic. For an arbitrary initial success probability, the rotation of the quantum state toward the target subspace does not, in general, reach the target state exactly after an integer number of iterations. Consequently, the optimal iteration count must be approximated by an integer, resulting in a success probability that can remain below unity. Several approaches have been proposed to address this limitation giving deterministic or near deterministic results,  including fixed-point quantum search and phase-based constructions that modify the search dynamics to improve or guarantee the success behaviour \cite{grover2005fixedpoint,yoder2014fixedpoint,  long2001grover, apqaa, dgsro, dqs, toyama2008multiphase, li2007phase}. Thus, although quantum search provides a quadratic query advantage, achieving deterministic success without compromising this advantage remains an important consideration.

In this work, we adopt a different approach to obtain deterministic quantum search for problems in the NP class. Rather than modifying the phase structure of the Grover operator, we build upon the ancilla-assisted deterministic search construction \cite{mishra2025deterministic, mishra2026SCA}. In this approach, an auxiliary qubit is used to modify the effective initial success probability such that the corresponding amplitude-amplification rotation reaches the target subspace exactly after an integer number of iterations. The construction therefore preserves the standard form of the search operator while requiring at most one additional iteration compared with the corresponding probabilistic procedure.

Building on this idea, we propose a general oracle construction for NP problems. The polynomial-time verifier associated with an NP problem is first represented as a Boolean circuit, which is subsequently transformed into a reversible circuit and incorporated into an oracle. The reversible implementation allows the classical verification procedure to be embedded into a quantum circuit while preserving its computational functionality \cite{bennett1989time}. This provides a systematic pipeline from the polynomial-time verifier to the quantum oracle, thereby separating the problem-specific construction of the verifier from the general procedure used to realize the corresponding quantum search oracle.

To demonstrate the applicability of the proposed framework, we instantiate the general oracle construction for the $N$-SAT problem and combine it with the deterministic quantum search procedure. The resulting implementation follows the complete transformation from the Boolean verifier to its reversible representation and finally to the oracle, followed by the deterministic amplitude-amplification procedure. The construction of the SAT oracle is based on the corresponding Boolean formulation and its quantum implementation \cite{portilheiro2018grover}. Simulation results demonstrate that the proposed procedure identifies the satisfying assignment deterministically, with the target state obtained with probability one for the considered instance. At the same time, the number of oracle queries retains the quadratic quantum-search scaling, given by $O(\sqrt{N/M})$ for a search space of size $N$ containing $M$ target states, up to the additional constant query overhead of the deterministic construction. Hence, the proposed framework combines the quadratic query advantage of quantum search with deterministic success, while the overall exponential dependence on the size of the NP search space remains unchanged. The construction therefore provides a general procedure for incorporating deterministic quantum search into NP problems using the polynomial-time verifier as the problem-specific component.

The paper begins with a review of the verifier-based characterisation of NP problems and the
classical exhaustive baseline, followed by Grover's algorithm, amplitude
amplification, and the deterministic quantum search framework adopted in this
work (Section~\ref{sec:preliminaries}). Next, the generalised construction of
the quantum oracle is presented, in which the polynomial-time verifier is
represented as a Boolean circuit, transformed into a reversible circuit, and
incorporated into the phase-marking operation of the deterministic search
(Section~\ref{sec:oracle}). The framework is then applied to the 
$N$-SAT problem, where the complete circuit is constructed and simulated for a
four-variable 4-SAT instance (Section~\ref{sec:nsat}). The complexity of the
proposed construction is analysed in terms of query complexity, circuit depth,
and qubit requirements (Section~\ref{sec:complexity}). Lastly, the paper concludes with a summary of
its contributions (Section~\ref{sec:conclusion}).

\section{From Classical Exhaustive Search to Deterministic Quantum Search}
\label{sec:preliminaries}

This section establishes the computational framework underlying the proposed approach. We first introduce the verifier-based characterisation of NP problems and the resulting candidate search space. We then consider the classical exhaustive approach and its corresponding worst-case complexity. Next, we discuss the standard quantum search algorithms that form the basis of the proposed framework, including Grover’s algorithm and amplitude amplification. These provide the quadratic query advantage of quantum search, but the standard search procedure remains probabilistic. Finally, we describe the deterministic quantum search framework adopted in this work, which addresses this limitation while retaining the quadratic query scaling of quantum search.

\subsection{The Verifier Characterisation of NP Problems}

The framework developed in this work follows directly from the verifier-based
characterisation of NP problems \cite{sipser2012theory, arora2009computational, cook2000p}. A problem $\Pi$ belongs to the class NP if, for every instance
$x$ with a valid solution, there exists a candidate solution $y$ that can be verified in polynomial
time by a verifier $V$. The verifier accepts the candidate solution when

\begin{equation}
    V(x,y)=1.
\end{equation}

Thus, for a fixed problem instance, the verifier defines a Boolean validity
function over the space of candidate solutions. The task of finding a solution
can therefore be viewed as a search for an input $y$ for which the verifier
returns one. If the candidate solution is represented by $n$ variables, each
taking a value from a set of size $d$, the corresponding candidate space
contains

\begin{equation}
    N = d^n
\end{equation}

possible assignments. This formulation provides a common representation of NP problems, where the problem-specific component is the polynomial-time
verifier, while the subsequent search procedure operates over the resulting
candidate space.






A direct classical strategy evaluates the verifier $V(x,y)$ independently for
each candidate $y$ in the search space. Let $V_c$ denote the computational
cost of one verifier evaluation. In the worst case, all $N$ candidates may
need to be examined, resulting in a query complexity of $O(N)$ and a
computational cost of

\begin{equation}
    T_{\mathrm{classical}}
    = O(NV_c),
    \qquad
    V_c=O(\mathrm{poly}(n)).
\end{equation}

For $N=d^n$, this gives
$T_{\mathrm{classical}}=O(d^nV_c)
=O(d^n\mathrm{poly}(n))$.

\subsection{Quantum Search Approaches}

Quantum search approaches provide an alternative to classical exhaustive search by reducing the number of queries required to identify a valid candidate solution. Grover's algorithm is the fundamental quantum approach for
unstructured search and provides a quadratic improvement in query complexity
over classical exhaustive search \cite{grover1996fast,grover1997quantum}. The
amplitude amplification framework generalises this procedure to arbitrary
initial states while retaining the same quadratic query advantage
\cite{brassard2002quantum}. These approaches provide the
quantum-search foundation for the deterministic construction considered in
this work.

Grover's algorithm and amplitude amplification provide a quadratic reduction in query complexity for unstructured search, requiring \(O(\sqrt{N/M})\) oracle queries for \(M\) valid solutions \cite{boyer1998tight,brassard2002quantum}. While amplitude amplification extends Grover's procedure to arbitrary initial states, the resulting search is generally probabilistic and does not guarantee exact identification of a valid solution after an integer number of iterations.

This probabilistic behaviour is a limitation for deterministic quantum search, where the goal is to identify a valid candidate solution to the NP problem with certainty.

This limitation motivates the deterministic quantum search framework adopted
in this work. Rather than modifying the underlying Grover search mechanism,
the proposed approach builds upon an ancilla-assisted deterministic
construction \cite{mishra2025deterministic} and integrates it with the
verifier-based formulation of NP problems. The following subsection describes
this deterministic framework and its use for achieving certain success while
retaining the quadratic query scaling of quantum search.

\subsection{The Deterministic Quantum Search Framework}

The deterministic quantum search (DQS) framework adopted in this work
follows the construction proposed in \cite{mishra2025deterministic}.
It modifies the effective initial success probability through an
auxiliary qubit, allowing amplitude amplification to identify the
target subspace with unit probability after an integer number of
iterations, without changing the underlying search oracle.

Let $p$ denote the initial probability of obtaining a valid solution.
The corresponding amplitude-amplification angle is
\begin{equation}
    \theta=\sin^{-1}(\sqrt{p}),
\end{equation}
and the ideal iteration count is
\begin{equation}
    k=\frac{\pi}{4\theta}-\frac{1}{2}.
\end{equation}
Since $k$ is not necessarily an integer, standard amplitude
amplification may not achieve unit success probability after an
integer number of iterations. The DQS construction addresses this
by calibrating the initial success probability to an effective value
$p'\leq p$ for which the required iteration count is an integer.

When $k$ is not an integer, the calibrated iteration count is chosen as
\begin{equation}
    k'=\lfloor k\rfloor+1,
\end{equation}
and the corresponding effective success probability is
\begin{equation}
    p'=\sin^2\left(\frac{\pi}{4(k'+1/2)}\right).
    \label{eq:effective_success_probability}
\end{equation}
The calibration is implemented by preparing an auxiliary qubit in the
state
\begin{equation}
    |\psi_{\mathrm{aux}}\rangle
    =
    \cos\left(\frac{\phi}{2}\right)|0\rangle
    +
    \sin\left(\frac{\phi}{2}\right)|1\rangle,
\end{equation}
using an $R_y(\phi)$ rotation. Since the target subspace is associated
with the $|0\rangle$ component of the auxiliary qubit, the effective
initial success probability becomes
\begin{equation}
    p'=p\cos^2\left(\frac{\phi}{2}\right).
\end{equation}
Consequently, the required rotation angle is
\begin{equation}
    \phi=2\cos^{-1}\left(\sqrt{\frac{p'}{p}}\right).
\end{equation}
For $p>0.5$, the construction uses the special calibration $p'=0.25$,
which permits deterministic identification in a single modified
amplification iteration.

The calibration changes the iteration count by at most one relative
to the ideal count and therefore preserves the asymptotic query
complexity of amplitude amplification. For a search space of size
$N$ containing $M$ valid solutions, the DQS query complexity is
\begin{equation}
    Q_{\mathrm{DQS}}
    =
    O\left(\sqrt{\frac{N}{M}}\right).
\end{equation}
Thus, DQS achieves deterministic identification of a valid solution
while retaining the quadratic query scaling of quantum search. In
this work, the framework is combined with the verifier-based oracle
construction described in the following section to support
deterministic quantum search for NP problems.


\section{Solving NP Problems using Deterministic Quantum Search}
\label{sec:oracle}

The central component of the proposed framework is the construction of a
quantum oracle from the polynomial-time verifier associated with an NP
problem. This section presents the detailed construction of the proposed
approach and describes how the classical verification procedure is integrated
into the deterministic quantum search framework. The construction ensures that a single evaluation of the resulting quantum
verifier oracle can be implemented using polynomially many gates and
workspace qubits, consistent with the polynomial-time complexity of the
classical verifier function. This permits a direct comparison between the proposed quantum search
approach and classical exhaustive search in terms of verifier-oracle
queries. The proposed procedure reduces the query complexity from
$O(N)$ for classical exhaustive search to
$O\left(\sqrt{\frac{N}{M}}\right)$ for deterministic quantum search,
where $N$ is the search-space size and $M$ is the number of valid
solutions.

\subsection{Solving NP Problems}

The quantum approach begins by representing the complete candidate space in
a quantum register. Let $x$ denote a fixed instance of an NP problem and let
$y$ denote a candidate solution. If each of the $n$ variables can take a value
from a set of size $d$, the candidate space contains

\begin{equation}
    N=d^n
\label{eq:search_space}
\end{equation}

possible assignments. The corresponding quantum register is prepared in a
superposition over these candidates. For a uniform preparation, the initial
state is

\begin{equation}
    \ket{\Psi_0}
    =
    \Aprep\ket{0}^{\otimes q}
    =
    \frac{1}{\sqrt{N}}
    \sum_{y}\ket{y},
\label{eq:uniform_super}
\end{equation}

where the summation is over all valid encodings of the candidate space. \\ $\Aprep$ represents the state preparation operator that is used to set up the system in a uniform superposition that includes every feasible candidate solution. Applying a Hadamard gate to each qubit that represents the input variables makes it easy to construct $\Aprep$ for problems with just binary assignments to each variable. The complexity of implementing $\Aprep$ can vary depending on the valid variable assignment from $\mathcal{O}(n \log d)$ \cite{shukla2024efficient} to $\mathcal{O}(n * d)$  \cite{Encoding_book, Dataencoding-2}.


The validity of each candidate is described by the Boolean function

\begin{equation}
    f_x(y)\equiv V(x,y),
\label{eq:validity_function}
\end{equation}

where $f_x(y)=1$ if $y$ is a valid solution for the instance $x$, and
$f_x(y)=0$ otherwise. The role of the quantum oracle is to evaluate this
validity condition coherently over the candidate superposition. Rather than
examining the candidates individually, the oracle acts on the superposition
and marks every valid candidate in the same operation.

This $f_x(y)$ is incorporated as a quantum operator, $B_x$ in the form of a standard oracle for query based algorithm. This oracle operates as,


\begin{equation}
    B_x\ket{y}\ket{b}
    =
    \ket{y}\ket{b\oplus f_x(y)},
\label{eq:bitflip_oracle}
\end{equation}

where $|b\rangle$ is an arbitrary computational-basis state of the
target qubit, $y$ is a candidate solution, and $\oplus$ denotes
addition modulo two. The oracle \(B_x\) may also act on additional qubits, which are omitted from Eq.\ref{eq:bitflip_oracle} for simplicity, since the action on these qubits is the identity. The detailed working of the operator $B_x$ is described in \ref{Subsubsec:Construction of oracle}.

The deterministic quantum search framework adjusts the effective initial
success probability so that the corresponding search finds the
valid-solution subspace after an integer number of iterations.
The auxiliary qubit is initialized using the $R_y(\phi)$ gate as described in the deterministic quantum search framework \cite{mishra2025deterministic}. To achieve determinism, the bit-flip oracle $B_x$, is incorporated with the prescribed single-qubit gates applied on the auxiliary qubit into the search operation to implement the controlled oracle. This
produces the required phase marking operator $U_{ctrl}$,

\begin{equation}
    U_{ctrl} \big( \ket{y}
    \otimes
    \left(cos(\frac{\phi}{2})\ket{0}+sin(\frac{\phi}{2})\ket{1}\right)\big)
    \longrightarrow
    \ket{y}
    \otimes
    \left(
    cos(\frac{\phi}{2})(-1)^{f_x(y)}\ket{0}
    +
    sin(\frac{\phi}{2})\ket{1}
    \right).
\label{eq:phase_marking}
\end{equation}

For a valid candidate, the $|0\rangle$ component of the auxiliary
qubit acquires a phase factor of $-1$, while the $|1\rangle$
component remains unchanged. For an invalid candidate, neither
component acquires a phase factor. This
phase difference is not directly observable in the measurement probabilities.
Its purpose is to provide the interference required by the subsequent
diffusion operation.

The diffusion operator then increases the amplitude of the states marked by the oracle. Repeated application ($k$-times) of the oracle and the diffusion operator amplifies the amplitude associated with the valid-solution subspace. 



The resulting
search therefore follows the general sequence

\begin{equation}
\boxed{
\text{State preparation}
\;\longrightarrow\;
[\text{Verifier oracle}
\;\longrightarrow\;
\text{Diffusion}]^k
\;\longrightarrow\;
\text{Measurement}
}
\label{eq:general_quantum_flow}
\end{equation}

according to the deterministic schedule, followed by measurement of
the candidate register. The subsequent construction makes this flow
problem-independent by isolating the verifier as the only
problem-specific component.













\subsection{Construction of Oracle from Verifier}

\subsubsection{From the Verifier to a Boolean Circuit}

By definition of NP, the verifier $V(x,\cdot)$ operates in polynomial time.
For any fixed problem instance $x$, the corresponding verification procedure can
therefore be represented by a Boolean circuit $C_x$ whose size and depth are
polynomial in the length of the binary encoding of the  problem instance \cite{sipser2012theory}. Thus the circuit computes the validity function
$f_x(y)$ for any candidate solution $y$ in polynomial time and the corresponding Boolean circuit satisfies

\begin{equation}
    |C_x|=\mathrm{poly}(L),
     \label{eq:boolean_circuit_size}
\end{equation}

with polynomial circuit depth, where $|C_x|$ is the number of gates in the circuit and $L=|x|$ denotes the length of binary encoding of problem instance. The important point is that the size of the verification circuit is polynomial in the size of the instance and candidate solution, and does not scale with the total number \(N=d^n\) of candidate assignments.

The Boolean circuit therefore provides a compact representation of the
problem-specific verification procedure. The subsequent steps of the oracle construction are dependent on the Boolean circuit representing the verifier processed from the specific structure of the underlying NP problem.

\subsubsection{From Boolean Circuit to Reversible Circuit}

A general Boolean circuit is not directly suitable for quantum computation,
since quantum operations must be reversible. The Boolean verification circuit
$C_x$ is therefore embedded into a reversible circuit
$\widetilde{C}_x$ using standard reversible computation techniques
\cite{bennett1989time,bennett1973logical}.

Let the candidate solution $y$ be encoded in a register of

\begin{equation}
    q=n\lceil\log_2 d\rceil
\end{equation}

qubits. Let $a$ denote the number of workspace qubits required by the
reversible implementation of the verifier. The reversible circuit can be
constructed to act as

\begin{equation}
\ \widetilde{C}_x:
|y\rangle |0\rangle^{\otimes a}|0\rangle
\longrightarrow
|y\rangle |g_x(y)\rangle
|f_x(y)\rangle,
\end{equation}

where $|g_x(y)\rangle$ denotes the workspace state generated during the
reversible evaluation of the verifier, and $|b\rangle$ is the auxiliary qubit used as the target of
the verification operation.

The candidate register $|y\rangle$ remains unchanged, while the verifier
result is XORed into the auxiliary qubit. Since the verifier circuit has
polynomial size, the reversible circuit requires only polynomially many
workspace qubits and reversible gates. In particular,

\begin{equation}
    a=O(|C_x|)=O(\mathrm{poly}(n)),
\end{equation}

and

\begin{equation}
    |\widetilde{C}_x|=O(|C_x|) =O(\mathrm{poly}(n)).
\end{equation}

Thus, the reversible embedding of the verifier introduces only polynomial computational overhead while preserving the polynomial scaling of the verification procedure.

\subsubsection{Construction of the Bit-Flip Oracle}
\label{Subsubsec:Construction of oracle}

The reversible implementation of the verifier can be used to construct the required bit-flip oracle. Denote
this oracle by $B_x$. Its action on the candidate register and the auxiliary
qubit is

\begin{equation}
    B_x
    |y\rangle |b\rangle
    =
    |y\rangle |b\oplus f_x(y)\rangle,
\end{equation}

with the polynomial-size workspace of the reversible verifier included
implicitly.

The important feature of this construction is that the same auxiliary qubit introduced in the deterministic search construction also serves as the target qubit of the bit-flip oracle. The verifier
therefore writes its validity bit directly into the DQS auxiliary qubit.

The bit-flip oracle is converted into the required phase marking operation
through phase kickback. Let the auxiliary qubit be initialized as

\begin{equation}
    |\psi_{\mathrm{aux}}\rangle
    =
    \alpha|0\rangle+\beta|1\rangle,
\end{equation}

where the amplitudes are determined by the $R_y(\phi)$ rotation used in the
deterministic search framework. To condition the phase marking on the
auxiliary state $|0\rangle$, the bit-flip oracle is surrounded by Pauli-$X$
and Hadamard operations on the same auxiliary qubit. The Hadamard gates convert the bit-flip operation into a phase operation
on the target qubit, while the surrounding $X$ gates interchange the
$|0\rangle$ and $|1\rangle$ components, thereby shifting the phase
marking to the $|0\rangle$ component required by the deterministic
search construction. The resulting operation
is

\begin{equation}
\label{eq:uctrl}
    U_{\mathrm{ctrl}}
    =
    (I\otimes X)
    (I\otimes H)
    B_x
    (I\otimes H)
    (I\otimes X).
\end{equation}

To establish the action of the controlled phase-marking operation, we initialize the workspace register \(w\) and the verifier-output qubit \(f\) to \(|0\rangle_w\) and \(|0\rangle_f\), respectively. Since the workspace consists of \(a\) qubits, so,
\[
|0\rangle_w = |0\rangle^{\otimes a}.
\]
The initial Pauli-\(X\) and Hadamard operations transform the auxiliary state as
\[
HX(\alpha|0\rangle+\beta|1\rangle)
= \alpha|-\rangle+\beta|+\rangle,
\]
where
\[
|-\rangle = \frac{|0\rangle-|1\rangle}{\sqrt{2}},
\qquad
|+\rangle = \frac{|0\rangle+|1\rangle}{\sqrt{2}}.
\]

For a given candidate \(y\), the reversible verifier circuit \(\widetilde C_x\) computes the verifier output \(f_x(y)\) and stores the intermediate results in the workspace qubits. Denoting these intermediate values by \(g_x(y)\), the state evolves as
\[
|y\rangle|0\rangle_w|0\rangle_f
\left(\alpha|-\rangle+\beta|+\rangle\right)
\xrightarrow{\widetilde C_x}
|y\rangle|g_x(y)\rangle_w|f_x(y)\rangle_f
\left(\alpha|-\rangle+\beta|+\rangle\right).
\]

The CNOT operation, controlled by the verifier-output qubit, applies \(X^{f_x(y)}\) to the auxiliary qubit. Since
\[
X|-\rangle=-|-\rangle,
\qquad
X|+\rangle=|+\rangle,
\]
its action is
\[
\begin{aligned}
&|y\rangle|g_x(y)\rangle_w|f_x(y)\rangle_f
\left(\alpha|-\rangle+\beta|+\rangle\right)
\quad\longrightarrow
\quad|y\rangle|g_x(y)\rangle_w|f_x(y)\rangle_f
\left(\alpha(-1)^{f_x(y)}|-\rangle+\beta|+\rangle\right).
\end{aligned}
\]
Thus, the verifier-dependent phase is introduced only on the \(|-\rangle\) component of the auxiliary state, while the \(|+\rangle\) component remains unchanged.

The inverse reversible circuit \(\widetilde C_x^\dagger\) is then applied to remove the intermediate verification results. Since the CNOT leaves the candidate, workspace, and verifier-output registers unchanged, the inverse computation restores these registers to their initial states:
\[
|y\rangle|g_x(y)\rangle_w|f_x(y)\rangle_f
\left(\alpha(-1)^{f_x(y)}|-\rangle+\beta|+\rangle\right)
\xrightarrow{\widetilde C_x^\dagger}
|y\rangle|0\rangle_w|0\rangle_f
\left(\alpha(-1)^{f_x(y)}|-\rangle+\beta|+\rangle\right).
\]

The uncomputation
ensures that no residual information about the verification computation
remains entangled with the candidate register. This is necessary for the
interference and amplitude amplification used in subsequent iterations.

Finally, the Hadamard and Pauli-\(X\) operations transform the auxiliary qubit back to the computational basis. Using
\[
XH|-\rangle=|0\rangle,
\qquad
XH|+\rangle=|1\rangle,
\]
the resulting state becomes
\[
|y\rangle|0\rangle_w|0\rangle_f
\left(\alpha(-1)^{f_x(y)}|0\rangle+\beta|1\rangle\right).
\]
The construction therefore preserves the verifier-dependent phase while restoring the workspace and verifier-output registers. With these registers returned to their initial states, the effective operation on the candidate and auxiliary qubits is precisely \(U_{\mathrm{ctrl}}\), as defined in Eq. \ref{eq:uctrl}.

Applying this operation to a computational basis state of the candidate
register gives

\begin{equation}
\begin{aligned}
U_{\mathrm{ctrl}}
\left(
|y\rangle
\otimes
(\alpha|0\rangle+\beta|1\rangle)
\right)
=
|y\rangle
\otimes
\left(
\alpha(-1)^{f_x(y)}|0\rangle
+
\beta|1\rangle
\right).
\end{aligned}
\end{equation}

Hence, when $f_x(y)=1$, the $|0\rangle$ component of the auxiliary
qubit acquires a phase factor of $-1$, while the $|1\rangle$
component remains unchanged. When $f_x(y)=0$, neither component
acquires a phase. The operation
therefore implements the required controlled phase marking without modifying
the internal structure of the bit-flip oracle.

The construction treats the verifier oracle as a black-box operation:
the internal structure of $B_x$ is unchanged, while the DQS-specific
phase marking is implemented externally through the surrounding
operations on the auxiliary qubit.


The complete oracle construction can therefore be summarised as :

\begin{equation}
\boxed{
V(x,\cdot)
\rightarrow
C_x
\rightarrow
\ \widetilde{C}_x
\rightarrow
B_x
\rightarrow
U_{\mathrm{ctrl}}
}
\end{equation}

where the final operation provides the conditional phase marking required by
the deterministic quantum search procedure.

\begin{figure}[h]

\centering

\begin{tikzpicture}[
    smallgate/.style={draw, thick, minimum size=0.62cm, fill=red!8,
                      rounded corners=1pt},
    rygate/.style={draw, thick, minimum height=0.7cm, minimum width=1.05cm,
                   fill=blue!8, rounded corners=1pt},
    initgate/.style={draw, thick, align=center, fill=green!8, rounded corners=1pt,
                     minimum height=0.95cm, minimum width=1.45cm},
    oracle/.style={draw, thick, align=center, fill=green!8, rounded corners=1pt,
                   minimum height=2.35cm, minimum width=2.3cm},
    diffusion/.style={draw, thick, align=center, fill=blue!8, rounded corners=1pt,
                      minimum height=2.35cm, minimum width=1.6cm},
    wire/.style={thick},
    every node/.style={font=\small}
]

\def\yreg{2.0}
\def\yanc{0.0}

\def\xin{-0.3}
\def\xinit{1.7}
\def\xXa{4.15}
\def\xHa{4.9}
\def\xoracle{7.0}
\def\xHb{9.1}
\def\xXb{9.85}
\def\xdiff{12.5}
\def\xmeas{14.2}
\def\xend{15.1}

\def\vhw{1.15}
\def\vtop{2.45}
\def\vbot{-0.45}


\draw[wire] (\xin,\yreg) -- (\xend,\yreg);
\draw[wire] (\xin,\yanc) -- (\xend,\yanc);

\draw[thick] (0.45,\yreg-0.16) -- (0.75,\yreg+0.16);
\node at (0.50,\yreg+0.33) {\scriptsize $q$};


\node[left] at (\xin,\yreg) {$\ket{0}^{\otimes q}$};
\node[left] at (\xin,\yanc) {$\ket{0}$};


\node[initgate] (init) at (\xinit,\yreg)
    {Initialise\\operator $\Aprep$};


\node[rygate] (ry) at (\xinit,\yanc) {$R_y(\phi)$};


\node[smallgate] (Xa) at (\xXa,\yanc) {$X$};
\node[smallgate] (Ha) at (\xHa,\yanc) {$H$};


\draw[thick, fill=green!8, rounded corners=1pt]
     (\xoracle-\vhw,\vbot) rectangle
     (\xoracle+\vhw,\vtop);

\node at (\xoracle,1.55) {Reversible};
\node at (\xoracle,1.20) {bit-flip oracle};
\node at (\xoracle,0.82) {$B_x$};

\node[fill=green!8, inner sep=1pt]
    at (\xoracle,\yreg) {\scriptsize $y$};

\node[fill=green!8, inner sep=1pt]
    at (\xoracle,\yanc) {\scriptsize $b\oplus f_x(y)$};


\node[smallgate] (Hb) at (\xHb,\yanc) {$H$};
\node[smallgate] (Xb) at (\xXb,\yanc) {$X$};


\draw[thick, fill=blue!8, rounded corners=1pt]
     (\xdiff-0.8,\yanc-0.45) rectangle
     (\xdiff+0.8,\yreg+0.45);

\node[align=center] at (\xdiff,{(\yreg+\yanc)/2})
    {Diffusion\\operator};


\draw[thick, rounded corners=1pt]
    (\xmeas-0.3,\yreg-0.3) rectangle
    (\xmeas+0.3,\yreg+0.3);

\draw[thick]
    (\xmeas-0.17,\yreg-0.08)
    arc[start angle=200,end angle=-20,radius=0.18];

\draw[thick, -latex]
    (\xmeas,\yreg-0.08) -- (\xmeas+0.13,\yreg+0.16);


\draw[thick, decorate,
      decoration={calligraphic brace, amplitude=8pt, mirror}]
    (\xXa-0.45,-0.95) -- (\xdiff+0.8,-0.95)
    node[midway, below=9pt] {Repeat $k'$ times};

\end{tikzpicture}

\caption{Generalised deterministic quantum search circuit for NP
problems. The candidate register is prepared by $\Aprep$, while a single
auxiliary qubit is prepared by $R_y(\phi)$ and used for oracle phase marking
within the repeated oracle-diffusion operators.}

\label{fig:generic_circuit}

\end{figure}
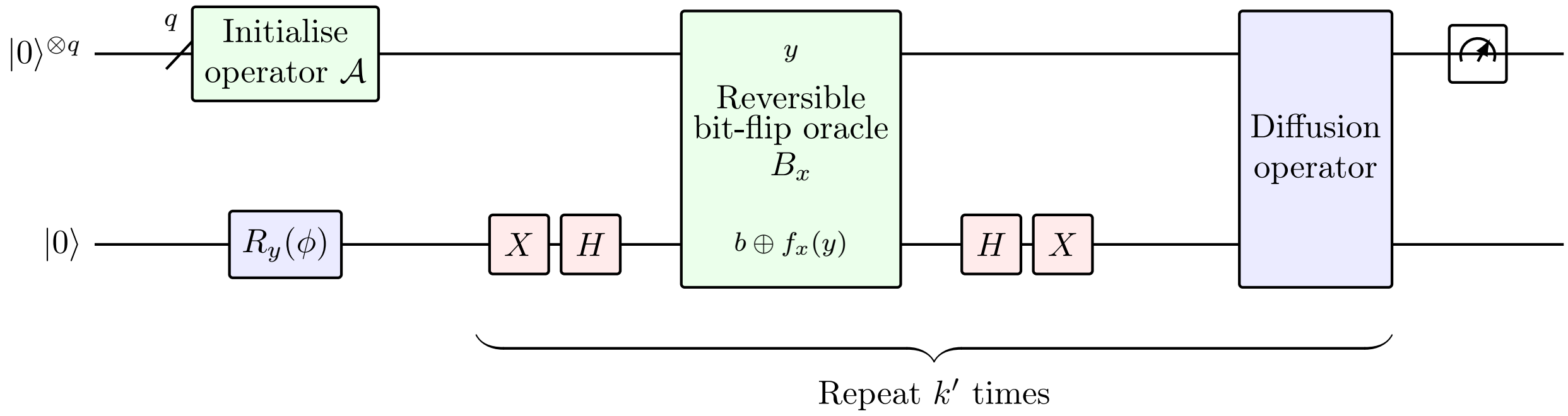

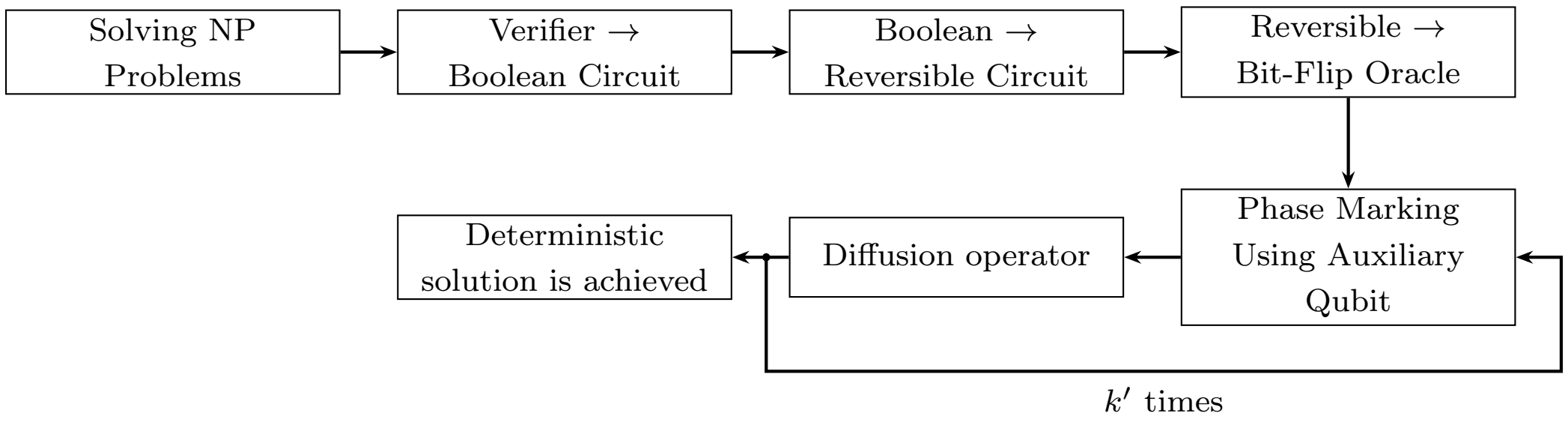
\begin{figure}[H]
\centering

\begin{tikzpicture}[
    block/.style={
        rectangle,
        draw,
        align=flush center,
        text width=0.17\columnwidth,
        minimum height=7mm,
        inner sep=2pt,
        font=\scriptsize
    },
    arrow/.style={
        -{Stealth[length=1.5mm]},
        thick
    },
    node distance=2mm and 5mm
]


\node[block] (np) {
    Solving NP\\
    Problems
};

\node[block, right=of np] (boolean) {
    Verifier $\rightarrow$\\
    Boolean Circuit
};

\node[block, right=of boolean] (reversible) {
    Boolean $\rightarrow$\\
    Reversible Circuit
};

\node[block, right=of reversible] (bitflip) {
    Reversible $\rightarrow$\\
    Bit-Flip Oracle
};


\node[block, below=8mm of bitflip] (phase) {
    Phase Marking\\
    Using Auxiliary Qubit
};

\node[block, left=of phase] (diffusion) {
    Diffusion operator
};

\node[block, left=of diffusion] (algorithm) {
    Deterministic \\
     solution is achieved
};


\draw[arrow] (np) -- (boolean);
\draw[arrow] (boolean) -- (reversible);
\draw[arrow] (reversible) -- (bitflip);


\draw[arrow] (bitflip) -- (phase);


\draw[arrow] (phase) -- (diffusion);
\draw[arrow] (diffusion) -- (algorithm);


\path (diffusion.west) -- (algorithm.east) coordinate[pos=0.4] (split);
\coordinate (loopbottom) at ([yshift=-4mm]phase.south);
\coordinate (loopright)  at ([xshift=4mm]phase.east);

\fill (split) circle (1pt);

\draw[arrow] (split)
    -- (split |- loopbottom)
    -- node[below, font=\scriptsize] {$k'$ times} (loopright |- loopbottom)
    -- (loopright)
    -- (phase.east);

\end{tikzpicture}

\caption{Block diagram of the proposed deterministic quantum search
framework for NP problems, showing the transformation from the
polynomial-time verifier to the quantum oracle and its integration into
the deterministic search procedure.}

\label{fig:np_dqs_framework}

\end{figure}

The proposed deterministic quantum search circuit is illustrated in
Figure~\ref{fig:generic_circuit}, while the overall framework and its
constituent steps are presented in the block diagram in
Figure~\ref{fig:np_dqs_framework}.

The flow of the proposed deterministic quantum search framework is summarised in Algorithm~\ref{alg:np_dqs}.

\begin{algorithm}[H]
\caption{Deterministic Quantum Search for NP Problems}
\label{alg:np_dqs}
\begin{algorithmic}[1]

\REQUIRE NP instance $x$, polynomial-time verifier $V$, $n$ variables,
domain size $d$, initial success probability $0<p<1$

\STATE Compute $N \gets d^n$.
\STATE Compute $k \gets \dfrac{\pi}{4\sin^{-1}(\sqrt{p})}-\dfrac{1}{2}$.

\IF{$k$ is an integer}
    \STATE Set $k' \gets k$ and $p' \gets p$.
\ELSE
    \STATE Set $k' \gets \lfloor k \rfloor+1$.
    \STATE Compute $p' \gets
    \sin^2\left(\dfrac{\pi}{4(k'+1/2)}\right)$.
\ENDIF

\STATE Compute $\phi \gets
2\cos^{-1}\left(\sqrt{\dfrac{p'}{p}}\right)$.
\STATE Prepare the candidate register using $\Aprep$.
\STATE Initialize the auxiliary qubit using $R_y(\phi)$.
\STATE Construct the Boolean circuit $C_x$ from $V(x,\cdot)$.
\STATE Convert $C_x$ into the reversible circuit $\widetilde{C}_x$.
\STATE Construct the bit-flip oracle $B_x$ from $\widetilde{C}_x$.
\STATE Construct $U_{\mathrm{ctrl}}$ using $X$ and $H$ operations
on the auxiliary qubit.

\FOR{$i=1$ to $k'$}
    \STATE Apply $U_{\mathrm{ctrl}}$.
    \STATE Apply the diffusion operator.
\ENDFOR

\STATE Measure the candidate register.

\ENSURE A valid solution $y$ with probability one.

\end{algorithmic}
\end{algorithm}

\section{Application of the Deterministic Quantum Search Framework to  N-SAT Instances}
\label{sec:nsat}

The generalised construction presented in
Section~\ref{sec:oracle} is illustrated through a detailed circuit-level implementation of a unique-solution 4-SAT instance.
The framework is further evaluated using additional 3-SAT and 4-SAT instances with different numbers of variables and satisfying
assignments, with the corresponding simulation results summarised in Table~\ref{tab:nsat_results}. The polynomial-time verifier evaluates the corresponding CNF formula to determine whether a candidate assignment satisfies all its clauses. This section presents the instance, the construction of its quantum oracle, and its
integration with the deterministic search procedure.
The circuit-level simulation is then used to
evaluate the identification of the unique satisfying
assignment.

















\subsection{The N-SAT Instance}

To demonstrate the proposed construction, we consider a 4-SAT
instance with four Boolean variables and a unique satisfying
assignment. For four variables, there are $2^4=16$ possible
four-literal clauses, each containing every variable either
in its original or negated form. Each such clause is falsified
by exactly one assignment. The instance is constructed by
including all possible clauses except the one falsified by
the desired satisfying assignment. Consequently, the resulting
formula contains fifteen clauses and has a unique satisfying
assignment. The CNF formula is

\begin{equation}
\begin{aligned}
F =\;&
(x_0\lor x_1\lor x_2\lor x_3)
\land
(x_0\lor x_1\lor x_2\lor\neg x_3)
\land
(x_0\lor x_1\lor\neg x_2\lor x_3)\\
&\land
(x_0\lor x_1\lor\neg x_2\lor\neg x_3)
\land
(x_0\lor\neg x_1\lor x_2\lor x_3)
\land
(x_0\lor\neg x_1\lor x_2\lor\neg x_3)\\
&\land
(x_0\lor\neg x_1\lor\neg x_2\lor x_3)
\land
(x_0\lor\neg x_1\lor\neg x_2\lor\neg x_3)
\land
(\neg x_0\lor x_1\lor x_2\lor x_3)\\
&\land
(\neg x_0\lor x_1\lor x_2\lor\neg x_3)
\land
(\neg x_0\lor x_1\lor\neg x_2\lor x_3)\\
&\land
(\neg x_0\lor\neg x_1\lor x_2\lor x_3)
\land
(\neg x_0\lor\neg x_1\lor x_2\lor\neg x_3)\\
&\land
(\neg x_0\lor\neg x_1\lor\neg x_2\lor x_3)
\land
(\neg x_0\lor\neg x_1\lor\neg x_2\lor\neg x_3).
\end{aligned}
\label{eq:nsat_formula}
\end{equation}





The parameters of the problem are,

\begin{equation}
    n=4,\qquad M=1,
\end{equation}

where $n$ is the number of variables, $M$ is
the number of satisfying assignments.

\subsection{Register Initialization and Deterministic Parameters}

Since the variables are Boolean, the candidate register contains $q=n=4$ qubits. The corresponding search space has $N=2^4=16$ candidate assignments. With one satisfying assignment, the initial success
probability is

\begin{equation}
    p=\frac{M}{N}=\frac{1}{16}.
\end{equation}

The ideal number of Grover-type iterations corresponding to this initial
probability is

\begin{equation}
    k=
    \frac{\pi}
    {4\sin^{-1}\sqrt{p}}
    -\frac{1}{2}.
\end{equation}

For $p=1/16$, this gives

\begin{equation}
    k\approx2.6083.
\end{equation}

Since the required number of iterations is not an integer, the deterministic
construction selects

\begin{equation}
    k'=\lfloor k\rfloor+1=3.
\end{equation}

The effective initial success probability is then chosen as in Eq.~\eqref{eq:effective_success_probability}, which gives

\begin{equation}
    p'
    =
    \sin^2\left(\frac{\pi}{14}\right)
    \approx0.0495.
\end{equation}

The auxiliary qubit is initialized using the rotation

\begin{equation}
    R_y(\phi)\ket{0}
    =
    \cos\left(\frac{\phi}{2}\right)\ket{0}
    +
    \sin\left(\frac{\phi}{2}\right)\ket{1},
\end{equation}

where

\begin{equation}
    \phi
    =
    2\cos^{-1}
    \sqrt{\frac{p'}{p}}
    \approx0.9465~\mathrm{rad}.
\end{equation}

The circuit therefore begins with four candidate qubits prepared by
$H^{\otimes4}$ and one auxiliary qubit prepared by $R_y(0.9465)$. In addition,
fifteen workspace qubits are allocated to store the intermediate satisfaction
results of the fifteen clauses.

The quantum oracle for the considered 4-SAT instance is
constructed by reversibly evaluating the individual
clauses and combining their satisfaction values to
determine the validity of each candidate assignment.
The resulting bit-flip oracle is incorporated into the
controlled phase-marking operation described in
Section~\ref{sec:oracle}. The complete reversible
circuit construction, including clause evaluation
and workspace uncomputation, is provided in ~\ref{app:nsat_oracle}.

\subsection{Circuit Implementation of the N-SAT Oracle and the Final Result}

The complete circuit implementation follows the sequence described above.
For the four-variable instance, the quantum circuit contains four candidate
qubits, fifteen clause workspace qubits, and one auxiliary qubit, giving a
total of

\begin{equation}
    Q=4+15+1=20
\end{equation}

 qubits. The fifteen clause qubits are used only as reversible workspace
for evaluating the individual clauses. The auxiliary qubit is shared between
the deterministic state preparation and the phase-marking operation, and no
separate oracle-output qubit is introduced.

The complete circuit was simulated for the four-variable 4-SAT instance using the
state vector simulator. After three deterministic oracle-diffusion iterations, the statevector simulation yields a candidate-register probability distribution in which the satisfying assignment ($1101$) has unit probability to numerical precision.

\begin{equation}
    x_3x_2x_1x_0=1101.
\end{equation}

which corresponds to

\begin{equation}
    x_0=1,\qquad
    x_1=0,\qquad
    x_2=1,\qquad
    x_3=1.
\end{equation}

Substitution of this assignment into Eq.~\eqref{eq:nsat_formula} satisfies all
fifteen clauses. Since the instance has a unique satisfying assignment, the
measured state is therefore the required solution.

The result obtained from the state vector simulation shows,

\begin{equation}
    P(1101)=1.
\end{equation}

\begin{figure}[H]
\centering
\includegraphics[width=0.6\textwidth]{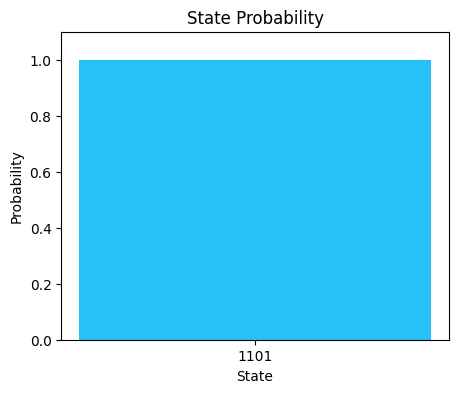}
\caption{Simulated probability for the four-variable 4-SAT instance after three deterministic search iterations. The state $\texttt{1101}$, corresponding to the unique satisfying assignment $x_3x_2x_1x_0=1101$, has unit probability.}
\label{fig:nsat4_result}
\end{figure}

Figure~\ref{fig:nsat4_result} shows the corresponding probability amplitude histogram.

The result demonstrates the deterministic behaviour of the proposed construction for the selected N-SAT instance. By calibrating the initial success probability through the auxiliary-qubit rotation, the framework enables the satisfying state to be reached with unit probability after three search iterations while retaining the quadratic oracle-query scaling.

Table~\ref{tab:nsat_results} summarises the state-vector simulation results for multiple N-SAT instances with varying numbers of variables
and satisfying assignments. For each instance, the auxiliary-qubit
rotation is calibrated according to the initial success probability.
The results show that the satisfying assignments are obtained with unit probability after the calibrated  oracle-diffusion iterations. For instances with two satisfying assignments, the final probability is equally distributed between the two satisfying states, with
each having a probability of $0.5$. These results demonstrate the deterministic behaviour of the proposed framework for the N-SAT instances.

\begin{table}[htbp]
\centering
\small
\begin{tabularx}{\textwidth}{@{}l c c c X@{}}
\toprule
\textbf{Instance} &
\shortstack{\textbf{Number of}\\\textbf{solutions}} &
\shortstack{\textbf{Required}\\\textbf{iterations}} &
\shortstack{\textbf{Ancillary rotation}\\\textbf{angle $\phi$ (rad)}} &
\textbf{Result(s)} \\
\midrule

3-SAT with 4 variables
& 1 & 3 & 0.9465
& $\ket{0111}$\\

3-SAT with 4 variables
& 2 & 2 & 1.0147
& $\ket{0000},\ket{1111}$ \\

3-SAT with 5 variables
& 2 & 3 & 0.9465
& $\ket{00000},\ket{11111}$ \\

4-SAT with 4 variables
& 1 & 3 & 0.9465
& $\ket{1101}$ \\

4-SAT with 4 variables
& 2 & 2 & 1.0147
& $\ket{0010},\ket{0111}$ \\

\bottomrule
\end{tabularx}
\caption{State-vector simulation results for multiple N-SAT instances
using the deterministic quantum search framework.}
\label{tab:nsat_results}
\end{table}

\section{Complexity Analysis}
\label{sec:complexity}

The computational resources of the proposed framework are determined by the
size of the candidate search space and the complexity of the corresponding
verifier. This section analyses the query complexity, circuit depth, and qubit
requirements of the construction. The cost of a single oracle--diffusion
iteration is considered separately from the total cost of the search.

\subsection{Query Complexity}

Let $N=d^n$ denote the size of the candidate search space and let $M$ denote
the number of valid solutions. A classical exhaustive approach may require
the verifier to be evaluated for each candidate individually. Its worst-case
query complexity is therefore

\begin{equation}
    Q_{\mathrm{classical}}=O(N).
\end{equation}

The proposed deterministic quantum search applies the verifier-based oracle
within the amplitude-amplification procedure. For $M$ valid solutions, the
number of oracle queries required by the deterministic construction is

\begin{equation}
    Q_{\mathrm{DQS}}
    =
    O\!\left(\sqrt{\frac{N}{M}}\right).
\end{equation}

For a unique solution, where $M=1$, this becomes

\begin{equation}
    Q_{\mathrm{DQS}}
    =
    O(\sqrt{N})
    =
    O\!\left(d^{n/2}\right).
\end{equation}

Thus, the proposed construction provides a quadratic improvement in query
complexity over classical exhaustive search. The deterministic calibration
changes the effective initial success probability so that the target subspace
is reached after an integer number of iterations, while preserving the
asymptotic query complexity of quantum search. The calibration introduces at
most one additional iteration relative to the corresponding
amplitude-amplification schedule, and therefore does not change the
asymptotic query complexity.

\subsection{Circuit Depth}

The depth of a single search iteration is determined primarily by the
reversible implementation of the verifier. For a fixed problem instance
$x$, the verifier $V(x,y)$ evaluates the candidate witness $y$.
Let $D_V(n)$ denote the depth of the corresponding Boolean verification
circuit $C_x$. Since the verifier operates in polynomial time, the depth
of its corresponding Boolean circuit is polynomial in the
input size \cite{sipser2012theory},

\begin{equation}
    D_V(n)=\mathrm{poly}(n).
\end{equation}

The Boolean verification circuit is converted into a reversible circuit
before being incorporated into the quantum oracle. This conversion introduces
only polynomial overhead, so the oracle circuit depth
$D_{\mathrm{oracle}}(n)$, which denotes the depth required for a single
evaluation of the reversible verification oracle, remains polynomial in $n$,

\begin{equation}
    D_{\mathrm{oracle}}(n)=\mathrm{poly}(n).
\end{equation}

The remaining operations in a search iteration, including the auxiliary-qubit
operations and the diffusion operator, also require polynomial depth with
respect to the number of variables. Hence, the circuit depth
$D_{\mathrm{iter}}(n)$, which denotes the depth of a single
oracle-diffusion iteration, can be expressed as

\begin{equation}
    D_{\mathrm{iter}}(n)
    =
    \mathrm{poly}\!\left(D_V(n),n\right)
    =
    \mathrm{poly}(n).
\end{equation}

The total circuit depth is obtained by applying the single-iteration
construction for $O(\sqrt{N/M})$ iterations. Therefore,

\begin{equation}
    D_{\mathrm{total}}
    =
    O\!\left(
    \sqrt{\frac{N}{M}}\,
    \mathrm{poly}(n)
    \right).
\end{equation}

Thus, the polynomial complexity of the verifier is preserved at the circuit
level for each search iteration, while the overall depth retains the
exponential dependence associated with the quantum search over the candidate
space.

\subsection{Qubit Requirements}

The qubit requirement of the proposed construction consists of the candidate
register, the reversible verification workspace, and the auxiliary qubit
used by the deterministic search procedure.

For a candidate solution in which each of the $n$ variables can take one of
$d$ possible values, the candidate register requires

\begin{equation}
    q=n\left\lceil\log_2 d\right\rceil
\end{equation}

qubits. Let $a$ denote the number of workspace qubits required by the
reversible implementation of the verifier. Since the Boolean
verification circuit has polynomial size, as established in
Eq.~\eqref{eq:boolean_circuit_size}, the reversible implementation
requires at most $O(|C_x|)$ workspace qubits,

\begin{equation}
    a=O(|C_x|)=\mathrm{poly}(n).
\end{equation}

The deterministic search construction requires one additional auxiliary
qubit. The total number of quantum qubits is therefore

\begin{equation}
    Q_{\mathrm{total}}
    =
    q+a+1.
\end{equation}

Substituting the corresponding bounds gives

\begin{equation}
    Q_{\mathrm{total}}
    =
    n\left\lceil\log_2 d\right\rceil
    +O(|C_x|)
    +1
    =
    O(\mathrm{poly}(n)).
\end{equation}

The auxiliary qubit is used both for the deterministic state preparation and
as the target of the bit-flip oracle. Consequently, no separate oracle-output
qubit is required. The workspace qubits used during reversible verification
are uncomputed after each oracle evaluation and therefore do not retain
intermediate verification information.


The proposed framework uses the ancilla assisted deterministic quantum search technique which requires the initial success probability to be known. In cases when the initial success probability is not known various amplitude estimation techniques \cite{brassard2002quantum, brassard1998quantum} can be used to know the amplitude of the target, thus calculating the number of iterations from there. Quantum search with an unknown number of solutions
has also been investigated in \cite{boyer1998tight}. However, incorporating such estimation procedures
would introduce additional computational overhead and is beyond the scope
of the present construction.


\section{Conclusion}
\label{sec:conclusion}

This work presented a general framework for applying deterministic quantum
search to problems in NP. The proposed approach is based on the
verifier-based formulation of NP problems, in which finding a solution is expressed as
a search over candidate inputs satisfying a polynomial-time verification
procedure. By converting the verifier into a Boolean circuit, transforming
the resulting circuit into a reversible implementation, and incorporating it
into a quantum oracle, the problem-specific verification procedure can be systematically integrated with the deterministic quantum search framework.
This construction separates the implementation of the verifier from the
search procedure and provides a systematic approach for applying deterministic
quantum search to different NP problems.

The proposed framework maintains the resulting query complexity of $O(\sqrt{N/M})$, compared with $O(N)$ for classical exhaustive search, where ($N$) is the size of the search space and ($M$) is the number of valid solutions. Furthermore, the reversible implementation of the polynomial-time verifier
requires only polynomially many gates and reusable workspace qubits with respect to the size of the binary encoding of the verifier function($L$) for each oracle
evaluation. Consequently, the overall gate complexity is $(O(\sqrt{N/M},{poly}(L)))$, where ($L$) denotes the size of the binary encoding of the verification function, provided the remaining operations in each iteration also have polynomial gate complexity in ($L$). This preserves the quadratic improvement in the number of search iterations while retaining polynomial gate complexity for the circuit implementation of the verifier.


The proposed framework was demonstrated for the N-SAT problem
through a detailed circuit-level implementation of a four-variable
4-SAT instance with a unique satisfying assignment. The circuit
uses reversible workspace qubits to evaluate the individual clauses
and employs the same auxiliary qubit for deterministic state
preparation and phase marking, without requiring any separate
oracle-output qubit. After three calibrated oracle-diffusion
iterations, the state-vector simulation confirmed that the unique
satisfying assignment was obtained with unit probability.
The framework was further evaluated on additional 3-SAT and
4-SAT instances with varying numbers of satisfying assignments. The corresponding simulations confirmed unit total success probability across all five considered instances, with the probability distributed among the satisfying assignments in cases involving multiple solutions. These results provide a circuit-level demonstration of deterministic success while retaining the quadratic
query scaling of quantum search.


Overall, this work establishes a systematic connection between classical polynomial-time verification and deterministic quantum search. We propose a generalised verifier-to-oracle construction and incorporate it into the deterministic amplitude amplification approach. This work demonstrates that the quadratic search advantage of quantum computation can be retained while replacing the probabilistic outcome of standard quantum search with deterministic
identification of a valid solution throughout the NP class problems.

\section*{Data Availability}

The code implementation of the proposed deterministic quantum
search framework for different 3-SAT and 4-SAT instances,
including cases with single and multiple satisfying assignments,
is publicly available in the GitHub repository:
\url{https://github.com/C-DAC-Bengaluru/Qniverse/tree/main/quantum-algorithm-developments/Deterministic\%20Quantum\%20Search\%20for\%20NP\%20problems}.

\section*{Acknowledgment}
    The quantum circuit implementations and corresponding results presented in this work were obtained using \textbf{Qniverse : A Unified Quantum Computing Platform} \cite{Qniverse}. This research was supported by the Ministry of Electronics and Information Technology (MeitY), Government of India, under the project titled \emph{“HPC-based Quantum Accelerators for Enabling Quantum Computing on Supercomputers”}, Grant No. 4(3)/2022-ITEA.

\appendix
\renewcommand{\thesection}{Appendix \Alph{section}}

\section{Reversible Construction of the N-SAT Oracle}
\label{app:nsat_oracle}

For the N-SAT instance considered in Section 4, the classical
verifier evaluates the CNF formula given in
Eq.~\eqref{eq:nsat_formula}. The corresponding quantum oracle
is constructed by evaluating each
clause separately and then combining the clause results to obtain the
satisfaction value of the complete formula \cite{portilheiro2018grover}.


Consider a clause

\begin{equation}
    C_i=l_{i1}\lor l_{i2}\lor\cdots\lor l_{i\ell},
\end{equation}

where each $l_{ij}$ is a literal. A clause is false only when all of its
literals are false. Hence, its falsifying condition can be written as

\begin{equation}
    \neg C_i
    =
    \neg l_{i1}
    \land
    \neg l_{i2}
    \land\cdots\land
    \neg l_{i\ell}.
\end{equation}

This form is convenient for reversible circuit construction because the
falsifying condition can be detected using a multi-controlled operation. For
each clause, the candidate qubits are transformed, when necessary, so that the
falsifying assignment corresponds to the all-$1$ control state. A
multi-controlled NOT operation then writes the clause result onto a dedicated
workspace qubit.

After the clause evaluation, the temporary transformations applied to the
candidate register are reversed. The workspace qubit is subsequently inverted
so that its computational-basis value represents the satisfaction of the
clause. Thus, after evaluating all $m$ clauses, the workspace register contains

\begin{equation}
    \ket{c_1c_2\cdots c_m},
\end{equation}

where $c_i=1$ indicates that clause $C_i$ is satisfied.

The complete CNF formula is satisfied only when all clause values are equal to
one. The global validity condition can therefore be evaluated using the clause
workspace as controls of a multi-controlled operation targeting the auxiliary
qubit. This produces the reversible bit-flip oracle

\begin{equation}
    B_x\ket{y}\ket{b}
    =
    \ket{y}\ket{b\oplus F(y)}.
\end{equation}

The clause workspace is not part of the candidate solution. It is temporary
workspace used to implement the polynomial-time verifier reversibly.

Following the construction of Section~\ref{sec:oracle}, the same auxiliary
qubit is then used for phase kickback. The bit-flip oracle is surrounded by the
$X$ and $H$ operations on the auxiliary qubit, giving

\begin{equation}
    U_{\mathrm{ctrl}}
    =
    (I\otimes X)
    (I\otimes H)
    B_x
    (I\otimes H)
    (I\otimes X).
\end{equation}

For the calibrated auxiliary state, this operation produces

\begin{equation}
\begin{aligned}
U_{\mathrm{ctrl}}
\left(
\ket{y}
\otimes
\left(
\cos\frac{\phi}{2}\ket{0}
+
\sin\frac{\phi}{2}\ket{1}
\right)
\right)
=
\ket{y}
\otimes
\left(
\cos\frac{\phi}{2}(-1)^{F(y)}\ket{0}
+
\sin\frac{\phi}{2}\ket{1}
\right).
\end{aligned}
\end{equation}

\begin{figure}[H]
    \centering
    \includegraphics[width=\textwidth]{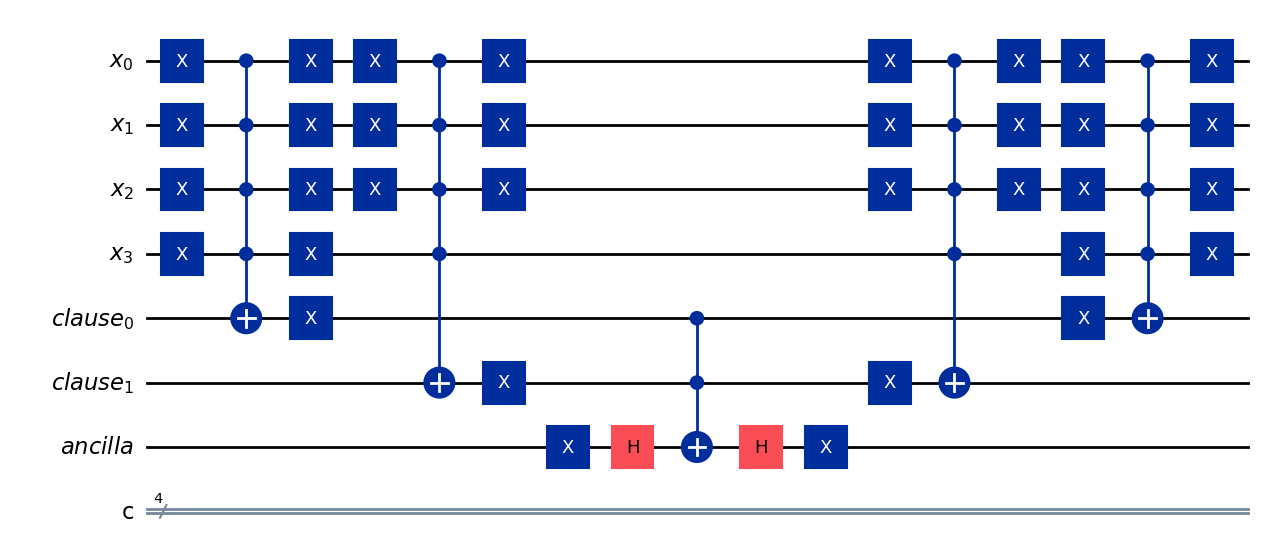}
    \caption{Quantum circuit for a single application of the proposed oracle to a two-clause 4-SAT instance, $F=(x_0\lor x_1\lor x_2\lor x_3)\land(x_0\lor x_1\lor x_2\lor\neg x_3)$ showing reversible clause evaluation, auxiliary-qubit phase marking, and workspace uncomputation.}
    \label{4-SAT oracle}
\end{figure}

Figure~\ref{4-SAT oracle} illustrates a single application of the quantum oracle constructed using the proposed approach for the two-clause CNF formula $F=(x_0\lor x_1\lor x_2\lor x_3)\land(x_0\lor x_1\lor x_2\lor\neg x_3)$, including reversible clause evaluation, auxiliary-qubit phase marking, and workspace uncomputation.

Therefore, satisfying assignments acquire the required phase inversion on the
$\ket{0}$ component of the auxiliary state. The clause workspace is then uncomputed by reversing the
clause-evaluation circuit, restoring all workspace qubits to their
initial states.

\printbibliography

\end{document}